\documentclass[a4paper,fleqn]{cas-sc}
\usepackage{soul}
\usepackage[sort&compress,numbers]{natbib}
\usepackage{lineno}
\usepackage{setspace}
\usepackage{verbatim}
\usepackage{graphics}
\usepackage{graphicx}
\usepackage{cuted}
\usepackage{lipsum}
\usepackage{multirow}

\def\tsc#1{\csdef{#1}{\textsc{\lowercase{#1}}\xspace}}
\tsc{WGM}
\tsc{QE}
\tsc{EP}
\tsc{PMS}
\tsc{BEC}
\tsc{DE}

\begin{document}

\let\WriteBookmarks\relax
\def\floatpagepagefraction{1}
\def\textpagefraction{.001}
\shorttitle{}
\shortauthors{K.A.L. Lima \textit{et~al}.}

\title [mode = title]{$\alpha$-, $\beta$-, and $\gamma$-TODD-G: Novel 2D Planar Carbon Allotropes}

\author[1,2]{Kleuton A. L. Lima}
\author[1,3]{Jose A. S. Laranjeiras}
\author[3]{Alysson M. A. Silva}
\author[4]{Bill. D. Aparicio-Huacarpuma }
\author[3]{Fabrício M. Vasconcelos}
\author[1,3]{Julio R. Sambrano}
\author[1,2]{Douglas S. Galvão}
\author[1,2]{Luiz A. Ribeiro Junior}
\cormark[1]
\ead{ribeirojr@unb.br}

\address[1]{Department of Applied Physics and Center for Computational Engineering and Sciences, State University of Campinas, 13083-859, Campinas, SP, Brazil}
\address[2]{Modeling and Molecular Simulation Group, School of Sciences, S\~ao Paulo State University, 17033-360, Bauru, SP, Brazil}
\address[3]{Department of Mechanical Engineering, College of Technology, University of Bras\'ilia, 70910-900, Bras\'ilia, DF, Brazil}
\address[4]{Institute of Physics, University of Bras\'ilia, 70910-900, Bras\'ilia, DF, Brazil}
\address[5]{Computational Materials Laboratory, LCCMat, Institute of Physics, University of Bras\'ilia, 70910-900, Bras\'ilia, DF, Brazil}
\address[6]{Instituto Federal do Piau\'i, IFPI, 64260-000, Piripiri, PI, Brazil}


\begin{abstract}
We report a comprehensive first-principles investigation of three novel two-dimensional carbon allotropes: $\alpha$-, $\beta$-, and $\gamma$-TODD-Graphene (TODD-G), composed of 3-8-12-16, 3-8-12-16, as 3-4-8-12 interconnected carbon rings with sp/sp$^2$ hybridization, respectively. Structural optimization, phonon spectra, and \textit{ab initio} molecular dynamics confirm their thermal and dynamic stability. All phases exhibit metallic electronic character, with distinct Dirac-like features and tilted Dirac cones suggesting anisotropic charge transport. Mechanical analysis reveals tunable anisotropy: $\alpha$-TODD-G is strongly anisotropic, $\beta$-TODD-G displays moderate anisotropy, and $\gamma$-TODD-G shows near-isotropic mechanical response. Optical spectra further distinguish the phases, with $\gamma$-TODD-G showing pronounced absorption in the infrared, while $\alpha$- and $\beta$-TODD-G absorb mainly in the visible and UV regions.
\end{abstract}




\maketitle
\doublespacing

Two-dimensional (2D) materials have emerged as one of the central focuses in materials science due to their distinct physical properties, derived from quantum confinement and reduced dimensionality \cite{dubertret2015rise,liu20192d}. Their planar nature allows for unusual mechanical flexibility, high surface area to volume ratios, and remarkable electronic and thermal transport characteristics \cite{xu2013graphene}. Among these, carbon-based 2D systems stand out due to the versatility of carbon bonds, which allows for a wide range of structural configurations and functional behaviors \cite{jana2021emerging}.

The potential applications of 2D materials span several domains, such as nanoelectronics, optoelectronics, catalysis, and energy storage \cite{zhu2018structural,khan2017two}. In particular, porous 2D carbon materials with adjustable pore sizes and multiple ring architectures have shown promising performance in lithium-ion and sodium batteries \cite{zhu2020ultrahigh,wang2013nitrogen}. Their intrinsic porosity favors ion mobility and adsorption capacity, resulting in better storage performance, while their atomic thickness allows for ultra-fast charge/discharge cycles. These properties also make them viable candidates for sensor technologies and flat optoelectronic devices \cite{zhao2023emerging,zheng2015two}.

In this context, a growing number of new 2D carbon allotropes have been proposed to overcome the limitations of graphene \cite{geim2009graphene}, especially the absence of an intrinsic bandgap \cite{tiwari2016magical,jana2021emerging}. Structures such as the biphenylene network \cite{fan2021biphenylene}, monolayer amorphous carbon \cite{toh2020synthesis,bai2024nitrogen}, holey graphene \cite{liu2020holey}, monolayer fullerene network \cite{hou2022synthesis,meirzadeh2023few}, graphyne \cite{desyatkin2022scalable,aliev2025planar}, and graphdiyne \cite{li2010architecture} were successfully synthesized through different experimental techniques. Moreover, several computationally predicted 2D carbon allotropes have also been designed. Examples are: popgraphene \cite{wang2018popgraphene}, phagraphene \cite{wang2015phagraphene}, $\psi$-graphene \cite{li2017psi}, T-Carbon \cite{sheng2011t}, DOTT-Carbon \cite{lima2025first}, PolyPyGY \cite{lima2025structural}, Petal-Graphyne \cite{lima2025petal}, Sun-Graphyne \cite{tromer2023mechanical}, Irida-Graphene \cite{junior2023irida}, and porous 2D anthraphenylenes \cite{lima2025anthraphenylenes}. Particularly, TOOD-Graphene \cite{santos2024proposing}, a 2D porous carbon allotrope composed of 3-8-10-12 carbon rings, was designed for superior lithium-ion battery efficiency. Its unique lattice topology enables easy structural tunability, allowing the proposal of novel stable variants with larger pore diameters.

In this work, we carried out a systematic computational characterization of three new porous 2D carbon allotropes, called $\alpha$-TODD-Graphene, $\beta$-TODD-Graphene, and $\gamma$-TODD-Graphene, as variants with larger pore diameters of TODD-Graphene. These materials have complex periodic architectures composed of 3-8-12-16, 3-8-12-16, as 3-4-8-12 interconnected carbon rings with sp/sp$^2$ hybridization, respectively. Using first-principles calculations based on density functional theory (DFT), we evaluated their structural and dynamical stability using formation energy, phonon scattering, and \textit{ab initio} molecular dynamics (AIMD) calculations. In addition, we calculate mechanical (elastic constants, in-plane Young's modulus), optical (dielectric function, absorption spectra), and electronic (band structure, density of states) properties, providing a comprehensive understanding of these new structures and their potential in flat electronics.

\section{Methodology}

All calculations were carried out within the framework of DFT using the CASTEP code \cite{clark2005first}. We employed norm-conserving pseudopotentials in conjunction with the Generalized Gradient Approximation (GGA) using the Perdew-Burke-Ernzerhof (PBE) exchange-correlation functional \cite{perdew1996generalized}. Structural relaxations were performed using the Broyden–Fletcher–Goldfarb–Shanno (BFGS) algorithm \cite{head1985broyden,PFROMMER1997233}, ensuring convergence of total energy to within $1.0\times10^{-5}$ eV, maximum residual forces below $1.0\times10^{-3}$ eV/\r{A}, and residual stress below $1.0\times10^{-2}$ GPa.

A vacuum layer of 20~\r{A} was included along the out-of-plane direction to eliminate spurious interactions between periodic images. The geometry optimization used a Monkhorst-Pack k-point mesh of $5\times5\times1$. For the calculation of the electronic band structure, density of states, and optical properties, denser meshes of $20\times20\times1$ were adopted. Optical properties were computed following the methodology detailed in reference \cite{lima2023dft}.

Phonon dispersion relations were obtained using density functional perturbation theory (DFPT) with a reciprocal space sampling of $5\times5\times1$, employing a finite displacement grid of 0.05~\r{A}$^{-1}$ and force convergence threshold of $1.0\times10^{-5}$ eV/\r{A}$^{2}$. Thermal stability was further validated through AIMD simulations at 1000~K in the NVT ensemble.

\section{Results}

\subsection{Structural properties}

\begin{figure}[pos=ht]
    \centering
    \includegraphics[width=\textwidth]{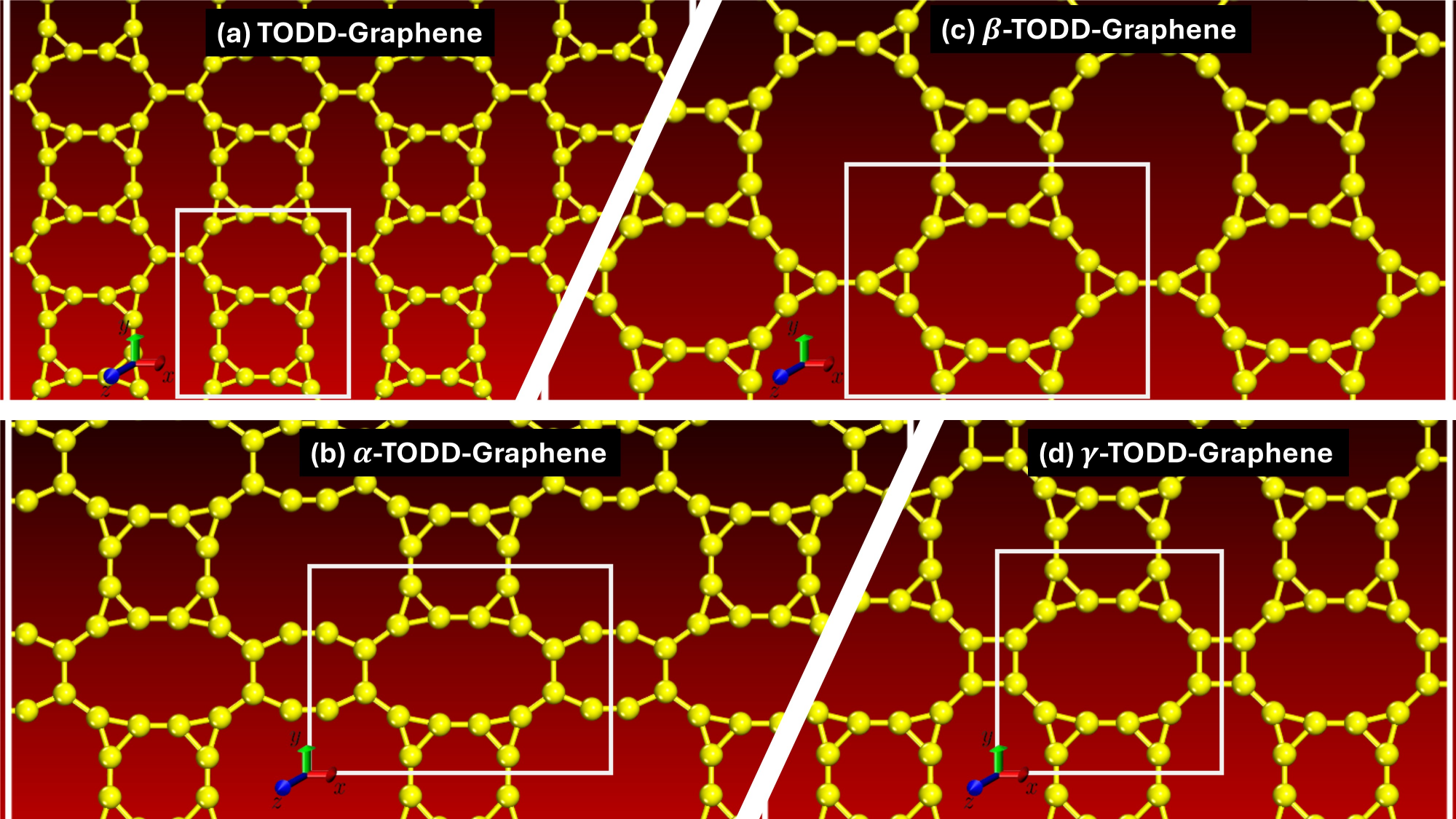}
    \caption{Top views of the atomic configurations of (a) TODD-Graphene, (b) $\alpha$-TODD-Graphene, (c) $\beta$-TODD-Graphene, and (d) $\gamma$-TODD-Graphene. The white boxes highlight the unit cells, and the Cartesian axes indicate the in-plane orientation. Yellow spheres represent carbon atoms.}
    \label{fig:structure}
\end{figure}

Figure~\ref{fig:structure} illustrates the fully relaxed atomic structures of TODD-Graphene (TODD-G) and its three derived allotropes, $\alpha$-, $\beta$-, and $\gamma$-TODD-G. All configurations exhibit a planar 2D topology. In this figure, white rectangles mark each unit cell for visual clarity. In this way, Figure \ref{fig:structure}(a) shows the original TODD-G structure, characterized by a regular pattern of 8- and 3-membered carbon rings aligned along the $x$ and $y$ directions, with lattice constants of $a = 7.03$~\AA\ ($x$-direction) and $b = 6.54$~\AA\ ($y$-direction). This framework serves as the foundational geometry for the other variants.

In $\alpha$-TODD-G (Figure~\ref{fig:structure}(b)), the introduction of additional carbon chains elongates the unit cell along the $x$-axis ($a = 10.20$~\AA), resulting in a more open and porous network. The $b$ parameter increases slightly to $7.13$~\AA, preserving orthorhombic symmetry. Figure \ref{fig:structure}(c) displays the $\beta$-TODD-G configuration, which retains the overall ring motif of the original TODD-G structure but introduces topological rearrangements that compress the unit cell along the $x$-axis ($a = 9.62$~\AA) while expanding it along the $y$-axis ($b = 7.66$~\AA). The $\gamma$-TODD-G structure, shown in Figure \ref{fig:structure}(d), features the most compact and symmetric arrangement among the three derivatives, with nearly equal in-plane parameters ($a = 7.54$~\AA, $b = 7.46$~\AA). The cohesive energy (E$_{coh}$) was calculated for $\alpha$, $\beta$, and $\gamma$-TODD-G, obtaining the values of -8.04 eV/atom, -7.98 eV/atom, and -7.94 eV/atom, respectively. These values demonstrate that the monolayers are energetically stable than the constituent ground state carbon atoms. 

\begin{figure}[pos=ht]
   \centering
    \includegraphics[width=\textwidth]{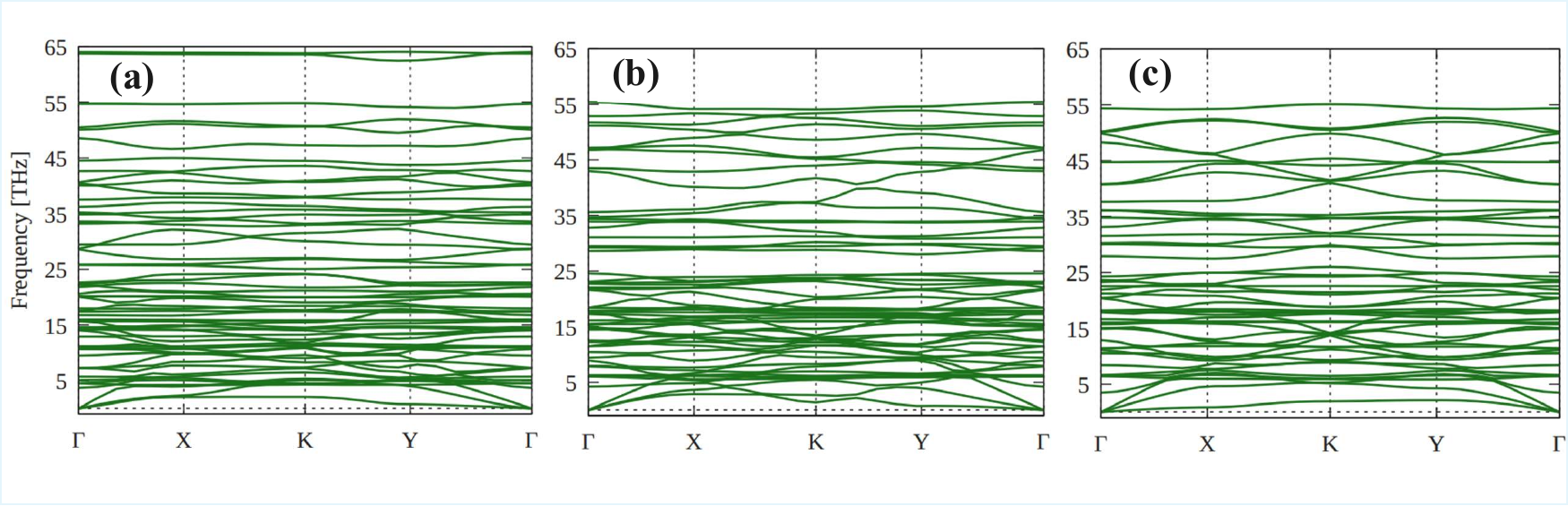}
    \caption{Phonon dispersion relations for (a) $\alpha$-TODD-G, (b) $\beta$-TODD-G, and (c) $\gamma$-TODD-G. No imaginary frequencies are observed across the Brillouin zone, confirming the dynamic stability of all structures.}
    \label{fig:phonon}
\end{figure}

Figure~\ref{fig:phonon} displays the phonon dispersion curves for the $\alpha$-, $\beta$-, and $\gamma$-TODD-G allotropes along high-symmetry paths in the rectangular Brillouin zone. The absence of imaginary frequencies in all spectra confirms that these carbon frameworks are dynamically stable at 0~K, with no tendency for structural distortion under small perturbations.

The phonon spectra of the three TODD-G monolayers are presented in Figure \ref{fig:phonon}. Among them, $\alpha$-TODD-G exhibits the broadest vibrational range, extending up to around 65~THz. This higher frequency limit results from the presence of acetylene-like (sp-hybridized) carbon bonds, which are not found in the other two structures.

In contrast, both $\beta$-TODD-G and $\gamma$-TODD-G show narrower phonon spectra, with maximum frequencies reaching approximately 55~THz. This reduction in vibrational frequencies is related to the absence of acetylene bonds in these structures. All three monolayers present well-behaved acoustic branches with smooth dispersion near the $\Gamma$ point, confirming their mechanical stability and robustness.

\begin{figure}[pos=ht]
    \centering
    \includegraphics[width=\textwidth]{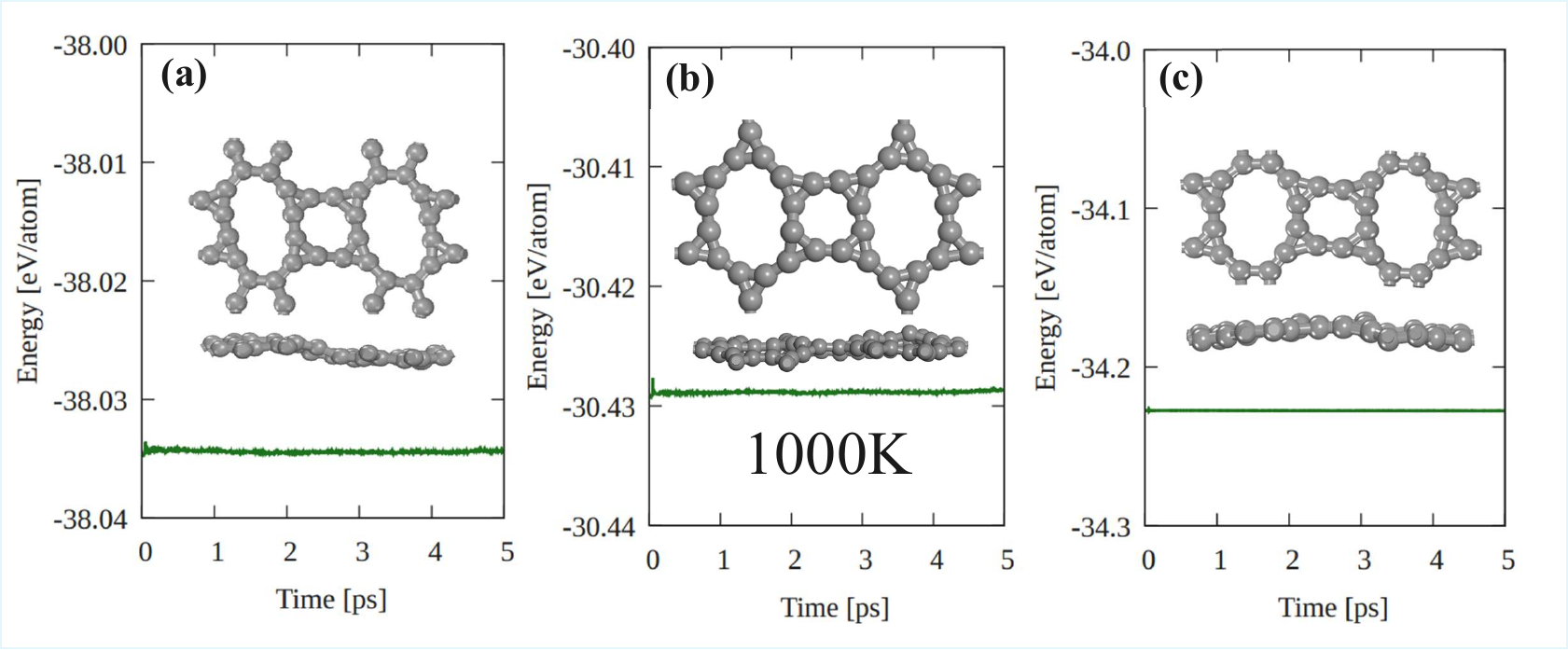}
    \caption{AIMD simulations at 1000~K for (a) $\alpha$-TODD-G, (b) $\beta$-TODD-G, and (c) $\gamma$-TODD-G. The inset panels show the top and side views of the final atomic configurations.}
    \label{fig:aimd}
\end{figure}

Figure~\ref{fig:aimd} presents the results of AIMD simulations performed at 1000~K for the three TODD-G allotropes over a 5~ps time window. Each panel displays the time evolution of the total energy per atom and the top and side views of the final atomic configuration, allowing for a direct assessment of thermal robustness and structural integrity under elevated temperature conditions.

Figure \ref{fig:aimd} shows the thermal stability of the three TODD-G monolayers at 1000~K. All structures ($\alpha$-, $\beta$-, and $\gamma$-TODD-G) maintain stable energy profiles throughout the simulation, with only minor thermal fluctuations and no signs of structural damage or bond rupture. The final configurations preserve their long-range order and porous topology, confirming the robust thermal and mechanical stability of these materials despite variations in anisotropy and pore geometry.

\subsection{Mechanical properties}

\begin{figure}[pos=ht]
    \centering
    \includegraphics[width=\textwidth]{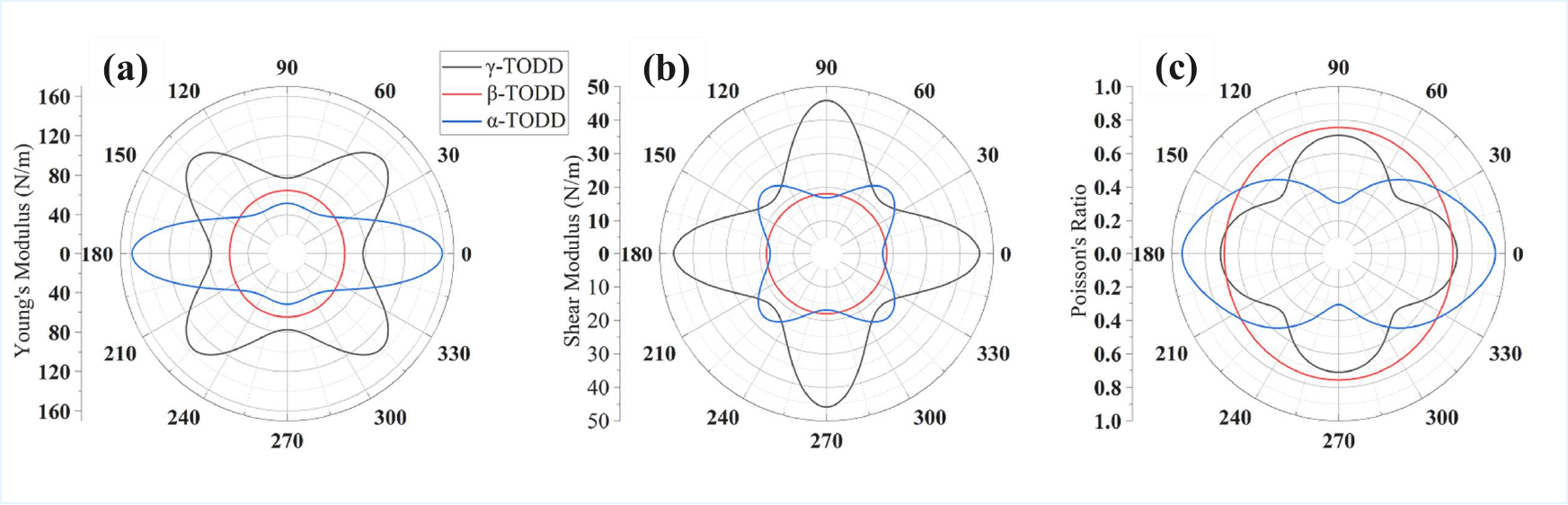}
    \caption{Polar plots of the (a) Young's modulus, (b) shear modulus, and (c) Poisson's ratio for $\alpha$-, $\beta$-, and $\gamma$-TODD-G.}
    \label{fig:Elastic}
\end{figure}

Figure~\ref{fig:Elastic} displays the in-plane mechanical response of $\alpha$-, $\beta$-, and $\gamma$-TODD-G derivatives through directional polar plots of Young's modulus ($E$), shear modulus ($G$), and Poisson's ratio ($\nu$). Table~\ref{tab:elastic_constants} summarizes the calculated elastic constants ($C_{ij}$) and the resulting mechanical properties—Young's modulus ($E$), shear modulus ($G$), and Poisson's ratio ($\nu$)—for the $\alpha$-, $\beta$-, and $\gamma$-TODD-G monolayers.

\begin{table}[pos=ht]
\centering
\caption{Elastic constants ($C_{ij}$ in N/m), Young's modulus ($E$ in N/m), shear modulus ($G$ in N/m), and Poisson's ratio ($\nu$) of $\alpha$-, $\beta$-, and $\gamma$-TODD-G monolayers.}
\label{tab:elastic_constants}
\begin{tabular}{lcccccccc}
\hline
\textbf{Material} & $C_{11}$ & $C_{22}$ & $C_{12}$ & $C_{66}$ & $E_{\text{min}}$ & $E_{\text{max}}$ & $G_{\text{max}}$ & $\nu_{\text{max}}$ \\
\hline
$\alpha$-TODD-G & 222.04 & 72.35 & 67.96 & 16.90 & 49.42 & 158.22 & 26.60 & 0.94 \\
$\beta$-TODD-G & 121.60 & 134.05 & 91.92 & 18.02 & 58.58 & 64.57 & 18.02 & 0.76 \\
$\gamma$-TODD-G & 155.69 & 155.69 & 110.45 & 45.83 & 77.34 & 136.36 & 45.83 & 0.71 \\
\hline
\end{tabular}
\end{table}

The elastic constants presented in Table~\ref{tab:elastic_constants} confirm the mechanical stability of all structures, as evidenced by the fulfillment of the Born–Huang stability criteria ($C_{11}>0$, $C_{66}>0$, and $C_{11}C_{22}>C_{12}^2$)~\cite{PhysRevB.90.224104,doi:10.1021/acs.jpcc.9b09593}.

Among the three monolayers, $\alpha$-TODD-G exhibits the highest anisotropy in Young's modulus, with values ranging significantly from 49.42~N/m to 158.22~N/m. In contrast, $\beta$-TODD-G displays almost isotropic mechanical behavior, as indicated by a narrow range of Young's modulus (58.58--64.57~N/m). The $\gamma$-TODD-G structure presents an intermediate scenario, with Young's modulus varying between 77.34 and 136.36~N/m.

Shear modulus values further reinforce these observations. $\gamma$-TODD-G has the highest maximum shear modulus (45.83~N/m), indicating superior resistance to shear deformation compared to the other structures. Conversely, $\alpha$- and $\beta$-TODD-G monolayers possess lower shear moduli (26.60 and 18.02~N/m, respectively), implying easier shear deformation, particularly in the $\beta$ allotrope.

Regarding Poisson's ratio, all three structures exhibit values within the physically stable range (0.3--0.9). $\alpha$-TODD-G has the highest maximum Poisson's ratio (0.94), suggesting significant lateral expansion when stretched longitudinally. In comparison, $\beta$- and $\gamma$-TODD-G possess lower maximum Poisson's ratios (0.76 and 0.71, respectively), indicating also higher (moderate when compared with $\alpha$-TODD-G) lateral expansion under applied uniaxial stress.


\subsection{Electronic properties}

\begin{figure}[pos=ht]
    \centering
    \includegraphics[width=\textwidth]{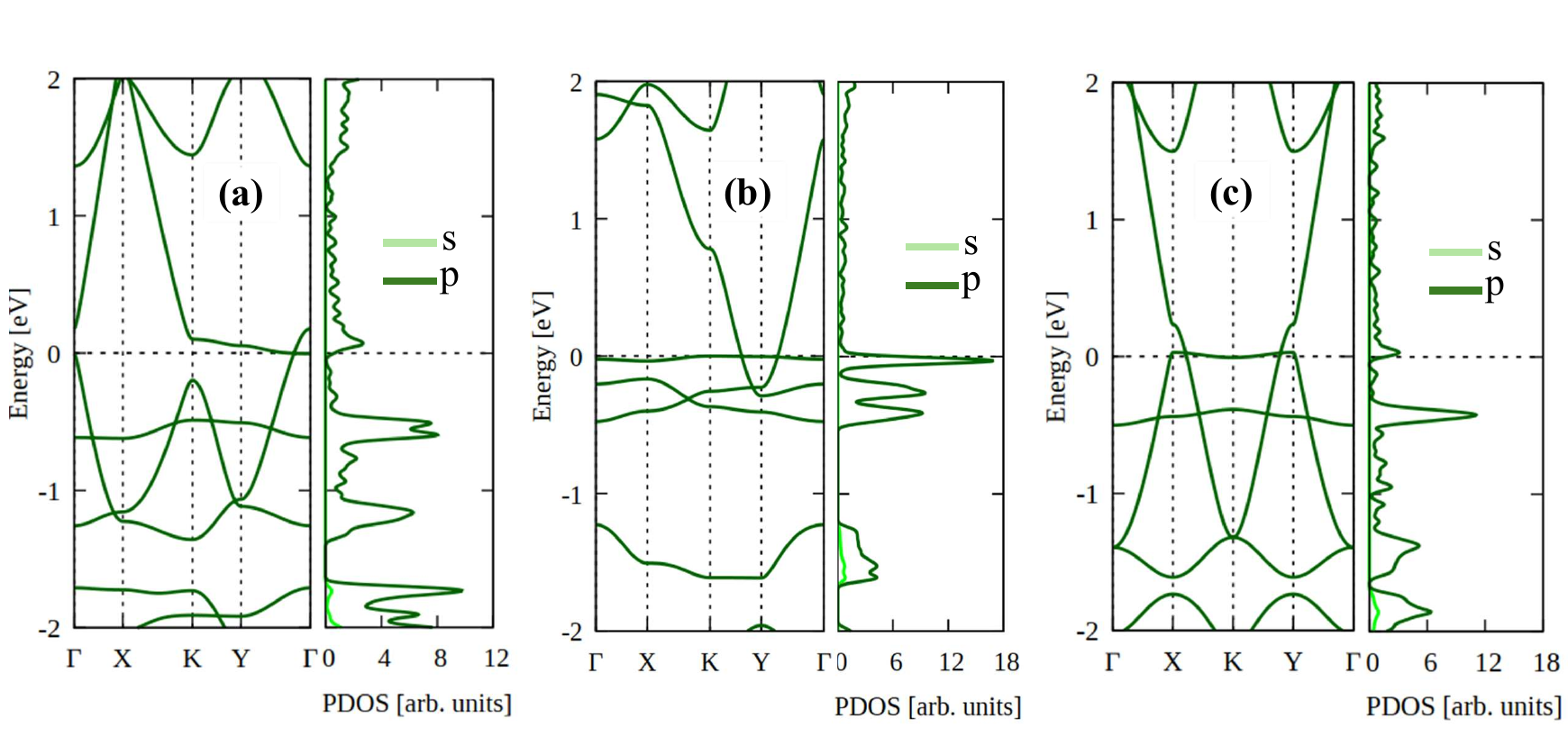}
    \caption{Electronic band structures and projected density of states (PDOS) for (a) $\alpha$-, (b) $\beta$-, and (c) $\gamma$-TODD-G. The Fermi level is set to 0 eV. Contributions from $s$ and $p$ orbitals are shown in light and dark green, respectively.}
    \label{fig:bands}
\end{figure}

Figure~\ref{fig:bands} presents the electronic band structures and projected density of states (PDOS) for the $\alpha$-, $\beta$-, and $\gamma$-TODD-G lattices. In all three cases, the electronic states cross the Fermi level, confirming the metallic nature of these materials. This behavior arises from the delocalized $\pi$-electron network formed by the $p$-orbitals, as highlighted in the PDOS, where the dominant contribution near the Fermi level originates from the $p$-type states (dark green lines).

Figure \ref{fig:bands} shows the electronic band structures and projected density of states (PDOS) for the three TODD-G allotropes. All three structures exhibit Dirac-like linear band crossings near the Fermi level, suggesting highly mobile massless fermion behavior predominantly driven by $p$-orbitals. Notably, $\alpha$-TODD-G presents moderately anisotropic linear bands, whereas $\beta$-TODD-G shows pronounced tilted Dirac cones along $K$–$Y$ and $X$–$\Gamma$ paths, reflecting stronger directional dependence in electronic transport. In contrast, $\gamma$-TODD-G features more symmetric and nearly ideal Dirac cones with minor tilt along $\Gamma$–$X$ and $Y$–$X$ directions, consistent with its more isotropic geometry and greater electronic delocalization, as also indicated by smoother PDOS profiles.

\begin{figure}[pos=ht]
    \centering
    \includegraphics[width=\textwidth]{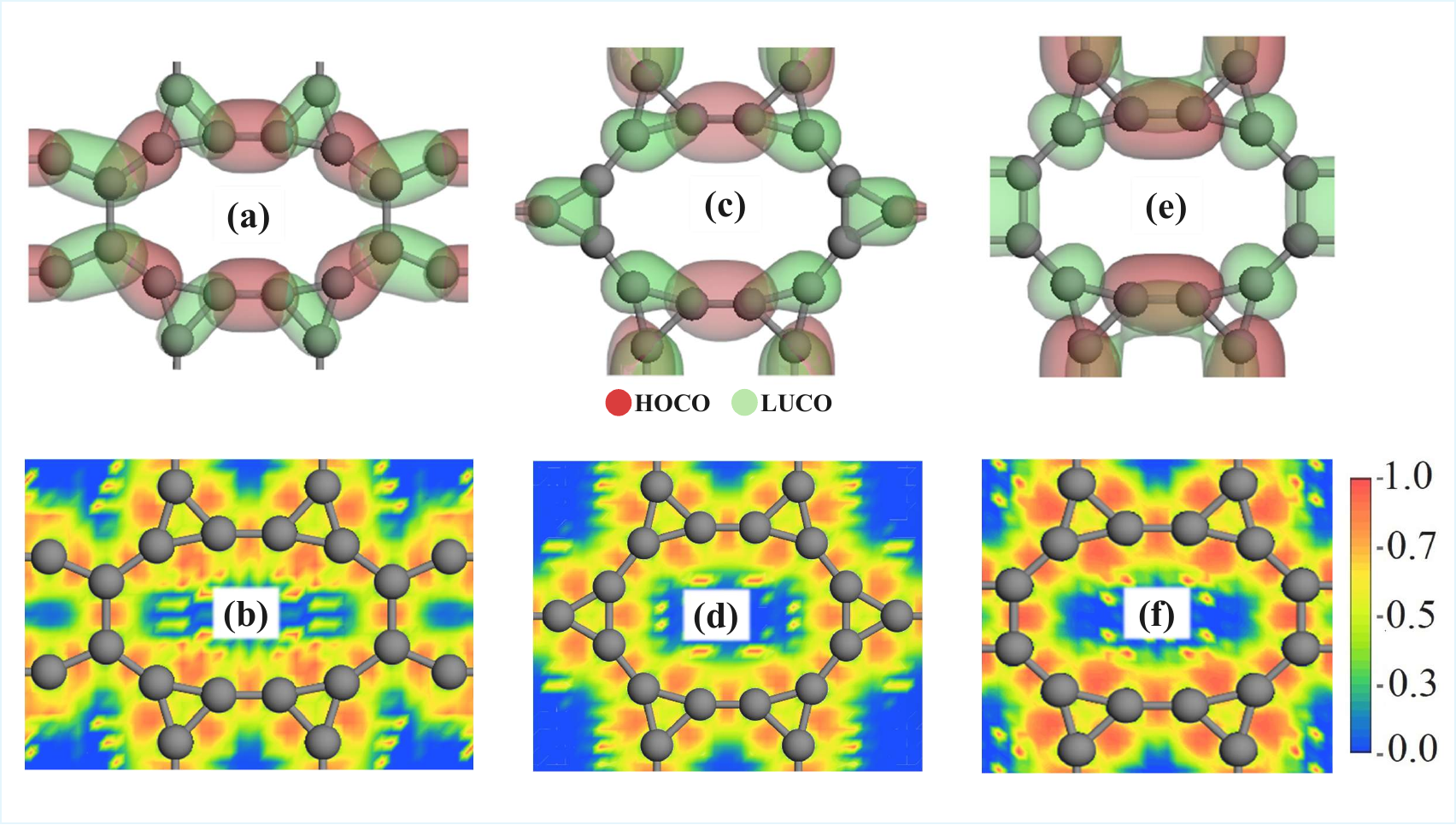}
    \caption{Frontier orbitals (HOCO in red and LUCO in green) and electron localization functions (ELF) for (a,b) $\alpha$-, (c,d) $\beta$-, and (e,f) $\gamma$-TODD-G. The ELF scale ranges from 0 (fully delocalized, blue) to 1 (fully localized, red).}
    \label{fig:orbitals_elf}
\end{figure}

Figure~\ref{fig:orbitals_elf} presents the spatial distributions of the highest occupied crystal orbital (HOCO), the lowest unoccupied crystal orbital (LUCO), and the corresponding electron localization function (ELF) maps for the three TODD-G phases. These visualizations provide insight into the nature of electronic delocalization, bonding character, and structural symmetry.

Figures \ref{fig:orbitals_elf}(a) and \ref{fig:orbitals_elf}(b) correspond to the $\alpha$-TODD-G structure. The frontier orbitals (HOCO in red and LUCO in green) are extensively delocalized across the lattice, particularly along the inter-ring bonds. This extended distribution is a hallmark of $\pi$-conjugation. The ELF map supports this picture, showing a relatively uniform distribution along the carbon–carbon bonds, with high localization values between atomic centers, consistent with strong covalent bonding.

The $\beta$-TODD-G derivative (Figures \ref{fig:orbitals_elf}(c) and \ref{fig:orbitals_elf}(d)) exhibits slightly higher localization for the frontier orbitals. Both HOCO and LUCO are concentrated around the central atoms of the larger carbon rings, indicating reduced delocalization and enhanced anisotropy in the electronic distribution. 

Figures \ref{fig:orbitals_elf}(e) and \ref{fig:orbitals_elf}(f) illustrate the electronic landscape of $\gamma$-TODD-G. Here, part of the LUCO orbitals are noticeably confined to 4-membered rings, with minimal overlap with HOCO. This spatial separation may be linked to lower charge carrier mobility and enhanced localization. The ELF map further confirms this behavior, displaying highly localized electron density in the ring cores and reduced bonding character along inter-ring connections. Despite this confinement, the structure remains stable, as demonstrated by its phonon and AIMD results.

\section{Optical properties}

\begin{figure}[pos=ht]
    \centering
    \includegraphics[width=\textwidth]{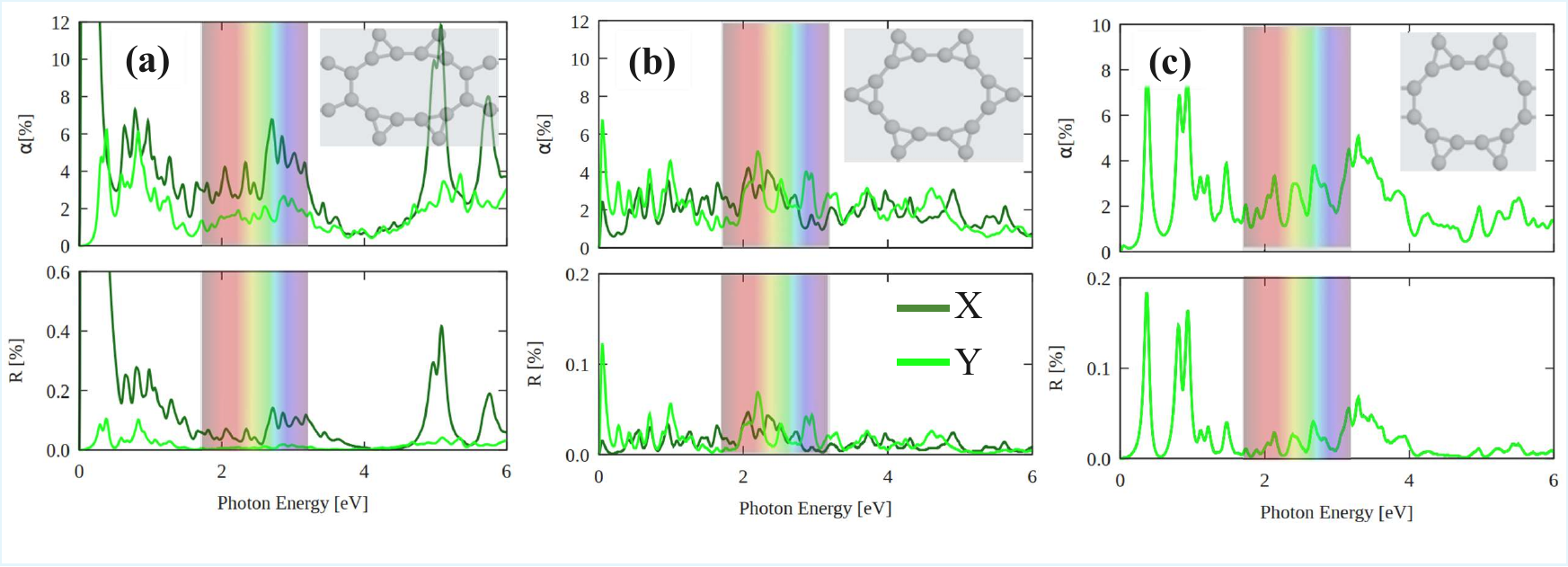}
     \caption{Optical absorption spectra and reflectivity for (a) $\alpha$-, (b) $\beta$-, and (c) $\gamma$-TODD-G along the $X$ (dark green) and $Y$ (light green) crystallographic directions.}
    \label{fig:optical}
\end{figure}

Finally, Figure~\ref{fig:optical} presents the optical absorption spectra and reflectivity curves for the three TODD-G analogues along the in-plane $x$ and $y$ directions. These results reveal distinct optical signatures for each material, highlighting their anisotropic light–matter interactions and their potential for optoelectronic applications across the infrared (IR), visible (VIS), and ultraviolet (UV) spectral regions.

The optical properties of $\alpha$-, $\beta$-, and $\gamma$-TODD-G are presented in Fig.~\ref{fig:optical}, where the absorption coefficient ($\alpha$) and reflectance ($R$) are shown for X- and Y-polarized light (dark and light green curves, respectively). All structures exhibit marked optical anisotropy, with distinct energy ranges for their main absorption peaks.

For $\alpha$-TODD-G (panel a), the X-component displays intense absorption peaks in the UV region (above 3.5~eV), while the Y-component shows significant features in the near-infrared (NIR) and visible region. This directional difference underscores its anisotropic response. The reflectance remains low throughout, below 0.6\%, with minor peaks aligning with strong absorption transitions.

In $\beta$-TODD-G (panel b), absorption is more moderate and spread across a broader energy range. In both components, the absorption onset occurs closer to 0 eV. The Y component presents a greater contribution to the absorption spectrum in the infrared. Multiple minor peaks are verified with relevant contributions in the visible range. The near-uniform reflectance, below 0.2\%, suggests limited backscattering and high optical transmittance. Some optical dichroism is verified for $\beta$-TODD-G; however, it is lower than that exhibited for the $\alpha$ counterpart.

For $\gamma$-TODD-G (panel c), the optical absorption and reflection occur isotropically, with the absorption dominated by a pronounced peak within the IR region (around 0.8~eV), showing several lower peaks for the visible range. The last significant absorption is visualized in the UVA; after this, the absorption decreases. Reflectance is minimal, assuming an analogous profile to the absorption, and is consistently below 0.2\%, similar to the other allotropes.

\section{Conclusions}

In summary, we performed a comprehensive first-principles investigation of three newly proposed two-dimensional carbon allotropes, named $\alpha$-, $\beta$-, and $\gamma$-TODD-Graphenel, characterized by hybrid ring architectures and mixed $sp/sp^2$ bonding networks. Structural optimization, phonon dispersion, and AIMD simulations confirmed the dynamical and thermal stability of all three phases. While $\alpha$- and $\beta$-TODD-G exhibit structural anisotropy due to elongated or asymmetric ring arrangements, $\gamma$-TODD-G shows a near-isotropic geometry, which reflects in its mechanical and optical behavior.

The mechanical response reveals tunable anisotropy across the allotropes: $\alpha$-TODD-G is strongly anisotropic with high stiffness along one principal axis; $\beta$-TODD-G exhibits moderate directional stiffness; and $\gamma$-TODD-G behaves nearly isotropically. Electronic band structures confirm metallic behavior for all three, with distinctive tilted Dirac-like features suggesting anisotropic massless carrier transport, especially in $\beta$- and $\gamma$-TODD-G. The spatial distribution of HOCO and LUCO orbitals, combined with ELF maps, shows progressive confinement of electronic density from $\alpha$ to $\gamma$, indicating reduced charge delocalization and transport efficiency in the latter.

Optically, each allotrope displays a distinct absorption signature: $\alpha$-TODD-G features strong UV-visible activity, $\beta$-TODD-G offers broadband low-reflectivity with modest visible absorption, and $\gamma$-TODD-G stands out for its infrared absorption. These findings highlight the versatility and tunability of the TODD-G family, suggesting their potential for application in flat nanoelectronics and novel boron nitride materials.

\section*{CRediT authorship contribution statement}
\noindent K.A.L.L., J.A.S.L., A.M.A.S., B.D.A.H., F.M.V.: Data curation, Formal analysis, Methodology, and Writing – Original draft preparation. J.R.S., D.S.G., and L.A.R.J.: Conceptualization, Funding acquisition, and Writing – Reviewing. All authors reviewed the manuscript. 

\section*{Declaration of competing interest}
\noindent The authors declare that they have no known competing financial interests or personal relationships that could have appeared to influence the work reported in this paper.

\section*{Data availability}
\noindent Data will be made available on request.

\section*{Acknowledgements}
The authors acknowledge the Molecular Simulation Laboratory at São Paulo State University (UNESP) and the Computational Materials Laboratory (LCCMat) at the University of Brasília for providing the computational facilities. 

\section*{Funding}
This work was supported by the Brazilian funding agencies Fundação de Amparo à Pesquisa do Estado de São Paulo (FAPESP) (grants no. 2022/03959-6, 2022/14576-0, 2013/08293-7, 2020/01144-0, 2024/05087-1, and 2022/16509-9), National Council for Scientific, Technological Development (CNPq) (grants no. 307213/2021–8, 350176/2022-1, and 167745/2023-9), FAP-DF (grants no. 00193.00001808/2022-71 and 00193-00001857/2023-95), FAPDF-PRONEM (grant no. 00193.00001247/2021-20), and PDPG-FAPDF-CAPES Centro-Oeste (grant no. 00193-00000867/2024-94). 

\bibliographystyle{unsrt}
\bibliography{cas-refs}
\end{document}